\documentstyle[12pt]{article}
\newcommand\be{\begin{equation}}
\newcommand\ee{\end{equation}}
\newcommand\bea{\begin{eqnarray}}
\newcommand\eea{\end{eqnarray}}
\newcommand\ket[1]{|#1\rangle}

\newcommand\braket[2]{\langle #1|#2\rangle}
\newcommand{\fatalpha}{{\bf \alpha \kern -0.44em \alpha}}
\newcommand{\fatsigma}{{\bf \sigma \kern -0.54em \sigma}}
\newcommand{\tpchi}{{\bf \chi \kern -0.35em \chi}}
\newcommand{\llambda}{{\bf \lambda \kern -0.45em \lambda}}

\renewcommand{\theequation}{\arabic{equation}}
\renewcommand{\theequation}{\thesection-\arabic{equation}}
\bibliography{plain}
\title{\bf Investigation of Continuous-Time Quantum Walk Via Spectral Distribution Associated
with Adjacency Matrix  }\vspace{20mm}
\author{ M. A. Jafarizadeh$^{a,b,c}$
 \thanks{E-mail:jafarizadeh@tabrizu.ac.ir}  ,
 S. Salimi$^{a,b}$
 \thanks{E-mail:shsalimi@tabrizu.ac.ir}
\\ $^a${\small Department of Theoretical Physics and Astrophysics,
Tabriz University, Tabriz 51664, Iran.} \\ $^b${\small Institute
for Studies in Theoretical Physics and Mathematics, Tehran
19395-1795, Iran.} \\ $^c${\small Research Institute for
Fundamental Sciences, Tabriz 51664, Iran.}} \pagebreak

\vspace{20mm}
\begin{document}
\maketitle \vspace{15mm}
\newpage
\begin{abstract}
Using the spectral distribution associated with the adjacency
matrix of graphs, we introduce a new method of calculation of
amplitudes of continuous-time quantum walk on some rather
important graphs, such as line,  cycle graph $C_n$, complete
graph $K_n$, graph $G_n$, finite path and some other finite and
infinite graphs, where all are connected with orthogonal
polynomials such as Hermite, Laguerre, Tchebichef and some other
orthogonal polynomials. It is shown that using the spectral
distribution, one can obtain the infinite time asymptotic
behavior of amplitudes simply by using the method of stationary
phase approximation(WKB approximation), where as an example, the
method is applied to star, two-dimensional comb lattices, infinite
Hermite and Laguerre graphs. Also by using the Gauss quadrature
formula one can approximate infinite graphs with finite ones and
vice versa, in order to derive  large time asymptotic behavior by
WKB method. Likewise, using this method, some new graphs are
introduced, where their amplitude are proportional to product of
amplitudes of some elementary graphs, even though the graphs
themselves are not the same as Cartesian product of their
elementary graphs. Finally, via calculating mean end to end
distance of some infinite graphs  at large enough times, it is
shown that continuous time quantum walk at different infinite
graphs belong to different universality classes which are also
different than those of the corresponding classical ones.


{\bf PACs Index: 03.65.Ud }
\end{abstract}
\vspace{70mm}
\newpage
\section{Introduction}
The theory of Markov chains and random walks on graphs is
fundamental to mathematics, physics, and computer science
\cite{d,ll,rmr} as it provides a beautiful mathematical framework
to study the stochastic process and its applications. Among  the
known examples of the applications we can mention Monte Carlo
methods in statistics, the theory of diffusion in statistical
physics, and algorithmic techniques for sampling and random
generation of combinatorial structures in computer science (based
on rapid mixing of certain Markov chains).
 Two pervasive algorithmic ideas in quantum computation are
Quantum Fourier Transform (QFT) and amplitude amplification (see
\cite{cn}). Most subsequent progress in quantum computing owed
much to these two beautiful ideas. But there are many problems
whose characteristics match neither the QFT nor the amplitude
amplification mold (e.g., the graph isomorphism problem), where
this begs for new additional tools to be discovered.

  A natural way to discover new quantum algorithmic ideas is to adapt
a classical one to the quantum model. An appealing well-studied
classical idea in statistics and computer science is the method
of random walks \cite{diaconis}. Recently, the quantum analogue
of classical random walks has been studied in a flurry of works
\cite{fg,cfg,abnvw,aakv,mr,k}. The works of Moore and Russell
\cite{mr} and Kempe \cite{k} showed faster bounds on
instantaneous mixing and hitting times for discrete and
continuous quantum walks on a hypercube (compared to the
classical walk).

A study of quantum walks on simple lattice is well known in
physics(see \cite{fls}). Recent studies of quantum walks on more
general graphs were described in \cite{fg,cfg,aakv,adz,ccdfgs}.
Some of these works studies the problem in the important context
of algorithmic problems on graphs and  suggests that quantum
walks is a promising algorithmic technique for designing future
quantum algorithms.

Several important classes of graphs studied in classical random
walks include the binary $n$-cube, the circulant graphs, and the
group-theoretic Cayley graphs. The binary $n$-cube and circulant
graphs are important in the study of interconnection networks and
complexity of Boolean function, and Cayley graphs capture strong
gorup-theortic ingredients of important problems, such as graph
isomorphism. Since most of these graphs are regular, the classical
random walks on them are known to converge or to mix towards the
uniform stationary distribution. The mixing properties of
continuous-time quantum walks on the same graphs were found to
exhibit non-classical behavior \cite{mr,abtw,aaht,gw}.

Here in this work we have used spectral distribution associated
with the adjacency matrix, of some particular graphs which possess
quantum distribution  ( for more details see Ref.\cite{nob}) in
studying continuous-time quantum walk. Spectral distribution
helps us to give a general expression for the amplitude of
observing the continuous-time quantum walker on a given site in
terms of the integral over product of the spectral distribution
and some polynomials (in most cases over the well-known orthogonal
polynomial; for the details see the appendix II). We use isometry
between the orthogonal polynomials (Hermite, Laguerre and
Tchebichef) and interacting Fock space associated with the
infinite graphs, representing raising (lowering) part adjacency
matrix, with first order differential operator, and obtain the
amplitude probability for infinite graphs associated to orthogonal
polynomials. In  obtaining the amplitude probability for  many of
the finite (complete $K_n$, complete cycle, $G_n$, finite path)or
infinite graphs(line, associated graphs of orthogonal polynomials
Hermite and Laguerre)  we also study the behavior of graph for
large vertices, and the asymptotic behavior of the walks. By using
direct product graphs, we obtain the new graphs and define
 an approach for calculating  the amplitude probability from the subgraphs.
  Also by using the Gauss quadrature formula one
can approximate infinite graphs with finite ones and vice versa,
which leads to the derivation of large time asymptotic
behavior\emph{continuous-time} quantum walk simply by using the
method of stationary phase. Also, we introduce graphs which have
different  structure,  but  they have uniform amplitude for
observing particle at  every level. At the end , via calculating
average visiting strata( sites for infinite line) which is
equivalent to mean end to end distance of some infinite graphs at
large enough times, it is shown that continuous time quantum walk
on infinite Hermite and line graphs belong to the same
universality class which is different from that of infinite
Laguerre one.

 The organization of this paper is as follows. In section
2, we give a brief outline of  graphs and their adjacency
matrices. In Section $3$,  we review the quantum decomposition
for adjacency matrix of graphs, and the method for obtaining
vacuum spectral distribution  $\mu$, and give some examples of
isometry from orthogonal polynomials into interacting Fock space.
Section $4$ is devoted to the method of computing amplitude
probability for continuous-time quantum walk, through spectral
distribution $\mu$ of the adjacency matrix $A$. Section $5$ is
concerned with  direct product of quantum decompositions for
finite graphs. In section $6$, we calculate  the amplitude
probability for continuous-time quantum walk on bunches of finite
and infinite graphs. In  section $7$, in first subsection we
evaluate the average visiting strata of some infinite graphs for
large enough time, and in remaining subsections,  large time
asymptotic behavior of quantum walks on different finite and
infinite graphs of section $6$ are obtained by using the method
of stationary phase(WKB) and Gauss quadrature rule. Paper is
ended with a brief conclusion and two appendices $I$ and $II$,
where the first appendix consists of the proof of lemma regarding
the equality of the amplitudes associated with vertices belonging
to the same stratum and the second appendix contains the list of
some of the orthogonal polynomials connected with some
particular  infinite and finite graphs, respectively.

\section{Graphs and its adjacency matrix}
In this section we give a brief outline of some of the main
features of graphs and walk on them, such as adjacency matrix,
stratification and orthonormal basis of strata. \\ A graph is a
pair $G=(V,E)$, where $V$ is a non-empty set and $E$ is a subset
of $\{(i,j);i,j\in V,i\neq j\}$. Elements of $V$ and of $E$ are
called \emph{vertices} and  \emph{edges}, respectively. Two
vertices $i; j\in V$ are called \emph{adjacent} if $(i,j)\in E$,
and in that case we write $i\sim j$. For a graph $G=(V,E)$ we
define the adjacency matrix $A=(A_{ij})_{i,j\in V}$ by

\[
A_{ij} = \left\{
\begin{array}{ll}
1 & \mbox{if $ i\sim j$}\\
0 & \mbox{otherwise.}
\end{array}
\right.
\]
Obviously, (i) $A$ is symmetric; (ii) an element of $A$ takes a
value in $\{0, 1\}$; (iii) a diagonal element of $A$ vanishes.
Conversely, for a non-empty set $V$, a graph structure is uniquely
determined by such a matrix indexed by $V$. The \emph{degree} or
\emph{valency} of a vertex $i\in V$ is defined by
$$
\kappa(i)=|\{j\in V; i\sim j\}|,
$$
where $\mid.\mid$ denotes the cardinality. A finite sequence $i_0;
i_1; . . . ; i_n \in V$ is called a walk of length $n$ (or of $n$
steps) if $i_{k-1}\sim i_k$ for all $k=1, 2, . . . , n$. In a walk
some vertices may occur repeatedly. Unless otherwise stated, we
always assume that a graph under discussion satisfies:\\
 (a) (connectedness) any pair of distinct vertices are connected by a
walk;\\
 (b) (local boundedness) $\kappa(i)< \infty$ for all $i\in
V$; In fact, the examples in this paper satisfy the following
condition which is stronger than (b):\\
$(b')$ (uniform boundedness) $sup_{i\in V} \kappa(i)<\infty$.\\
 Let $l^2(V)$ denote the Hilbert space of C-valued square-summable functions on
 V, and $\{\ket{i};i\in V\}$ becomes a complete orthonormal basis of
$l^2(V)$. The adjacency matrix is considered as an operator
acting in $l^2(V)$ in such a way that
$$
A\ket{i}=\sum_{i\sim j}\ket{j}, \;\;\;\;i\in V.
$$
Then, $A$ becomes a self-adjoint operator equipped with a natural
domain. As is easily checked, $(b')$ is a necessary and sufficient
condition for $A$ to be a bounded operator on $l^2(V)$.
\subsection{Stratification}
For $i\neq j$ let $\partial(i,j)$ be the length of the shortest
walk connecting $i$ and $j$. By definition $\partial(i,j)=0$ for
all $i\in V$. The graph becomes a metric space with the distance
function $\partial$. Note that $\partial(i,j)=1$ if and only if
$i\sim j$. We fix a point $o\in V$ as an origin of the graph.
Then, the graph is stratified into a disjoint union of strata:
\begin{equation}\label{v1}
V=\bigcup_{k=0}^{\infty}V_k,\;\;\;\;\;\; V_k=\{i\in
V;\partial(o,i)=k\}.
\end{equation}
With each stratum $V_k$ we associate a unit vector in $l^2(V)$
defined by
\begin{equation}
\ket{\phi_{k}}=\frac{1}{\sqrt{|V_k|}}\sum_{i\in V_{k}}\ket{k, i},
\end{equation}
where, $\ket{k, i}$ denotes the eigenket of $i$th vertex at the
stratum $k$.
 The closed subspace of $l^2(V)$ spanned by
$\{\ket{\phi_{k}}\}$ is denoted by $\Gamma(G)$. Since
$\{\ket{\phi_{k}}\}$ becomes a complete orthonormal basis of
$\Gamma(G)$, we often write
\begin{equation}
\Gamma(G)=\sum_{k}\oplus \textbf{C}\ket{\phi_{k}}.
\end{equation}
\section{Quantum decomposition}
In this section, first we review the quantum decomposition for
adjacency matrix of some particular graphs called QD graphs, and
the method for obtaining the vacuum spectral distribution  $\mu$
(for more detail see Ref. \cite{obah}), then we give  some
examples of isometry from orthogonal polynomials into interacting Fock space. \\
Let $A$ be the adjacency matrix of a graph $G=(V,E)$. According
to the stratification (\ref{v1}), we define three matrices $A^+$,
$ A^-$ and $A^0$ as follows: for $i\in V_k$ we set
\[
(A^+)_{ji} = \left\{
\begin{array}{ll}
A_{ji} & \mbox{if $ j\in V_{k+1}$}\\
0 & \mbox{otherwise.}
\end{array}
\right.
\]
\[
(A^-)_{ji} = \left\{
\begin{array}{ll}
A_{ji} & \mbox{if $ j\in V_{k-1}$}\\
0 & \mbox{otherwise.}
\end{array}
\right.
\]
\[
(A^0)_{ji} = \left\{
\begin{array}{ll}
A_{ji} & \mbox{if $ j\in V_{k}$}\\
0 & \mbox{otherwise.}
\end{array}
\right.
\]
Or equivalently, for $\ket{k,i}$,
\begin{equation}\label{qd}
A^{+}\ket{k,i}=\sum_{j\in V_{k+1}}\ket{k+1,j}, \;\;\;\;\ A^{-}\ket
{k,i}=\sum_{j\in V_{k-1}}\ket{k-1,j}, \;\;\;\;\
A^{0}\ket{k,i}=\sum_{j\in V_{k}}\ket{k,j}, \;\;\;\;\
 \end{equation}
 for $j\sim i$.
Since $i\in V_k$ and $i\sim j$ then $j\in V_{k-1}\bigcup
V_k\bigcup V_{k+1}$, where we tacitly understand that
$V_{-1}=\emptyset$, Now with the help of \cite{nob} we define
\begin{equation}
A=A^{+}+A^{-}+A^0.
\end{equation}
This is called  quantum decomposition of $A$ associated with the
stratification (\ref{v1}). Note also that
\begin{equation}
(A^+)^\star=A^- , \;\;\;\;\;\;\;\;\ (A^0)^\star=A^0,
\end{equation}
which can be verified easily. The vector state corresponding to
$\ket{o}=\ket{\phi_0}$, with $o\in V$ as the fixed origin, is
analogous to the vacuum state in Fock space.
 According to Ref.\cite{nob}, the
$<A^m>$ coincides with the number of $m$-step walks starting and
terminating at $o$, also, by \textbf{lemma 2.2}, \cite{nob} if
$\Gamma(G)$ is invariant under the quantum components
$A^\varepsilon$, $\varepsilon\in \{+,-,0\}$,  then there exist two
Szeg\"{o}- Jacobi sequences $\{\omega_k\}_{k=1}^{\infty}$ and
$\{\alpha_k\}_{k=1}^{\infty}$ derived from $A$, such that
\begin{equation}\label{v5}
A^{+}\ket{\phi_{k}}=\sqrt{\omega_{k+1}}\ket{\phi_{k+1}}, \;\;\;\
k\geq 0
\end{equation}
\begin{equation}\label{v6}
A^{-}\ket{\phi_{0}}=0, \;\;\
A^{-}\ket{\phi_{k}}=\sqrt{\omega_{k}}\ket{\phi_{k-1}}, \;\;\;\
k\geq 1
\end{equation}
\begin{equation}\label{v7}
A^{0}\ket{\phi_{k}}=\alpha_{k+1}\ket{\phi_{k}}, \;\;\;\ k\geq 0,
\end{equation}
where
$\sqrt{\omega_{k}}=\frac{|V_{k+1}|^{1/2}}{|V_{k}|^{1/2}}\kappa_{-(j)}$,
$\kappa_{-(j)}=|\{i\in V_k;i\sim j\}|$ for $j\in V_{k+1}$ and
$\alpha_{k+1}=\kappa_{0(j)}$, such that $\kappa_{0(j)}=|\{i\in
V_k;i\sim j\}|$ for $j\in V_k$.\\
Obviously from the relation between $\omega_k, \alpha_k,
\kappa_{-}(y)$ and $\mid V_k\mid$ it follows that $\mid
V_k\mid-1\geq \alpha_{k+1}$ and $\frac{\mid V_k\mid
\alpha_{k+1}}{2}$ must be integer.

From now on we call the graphs with the above property as  kids
of graphs which possess quantum distribution or QD type.

\section{Spectral distribution $\mu$ of the adjacency matrix $A$}
 The spectral properties of the adjacency matrix of a
graph play an important role in many branches of mathematics and
physics . The spectral distribution can be generalized in various
ways. In this work,  following Ref.\cite{nob}, we consider the
spectral distribution $\mu$ of the adjacency matrix $A$:
\begin{equation}\label{v2}
<A^m>=\int_{R}x^{m}\mu(dx), \;\;\;\;\ m=0,1,2,...
\end{equation}
where $<.>$ is the mean value with respect to a state
$\ket{\phi_0}$(the ket corresponds to the ground stratum). By the
local boundedness condition (b) of section $3$ the ''moment''
sequence
 $\{<A^m>\}_{m=0}^{\infty}$ is well-defined\cite{nob}. Then
the existence of a spectral distribution satisfying (\ref{v2}) is
a consequence of Hamburger's theorem, see e.g., Shohat and
Tamarkin
[\cite{st}, Theorem 1.2].\\
 We may apply the canonical isomorphism from
the interacting Fock space onto the closed linear span of the
orthogonal polynomials determined by the Szeg\"{o}-Jacobi
sequences $(\{\omega_k\},\{\alpha_k\})$. More precisely, the
spectral distribution $\mu$ under question is characterized by the
property of orthogonalizing the polynomials $\{Q_n\}$ defined
recurrently by
$$ Q_0(x)=1, \;\;\;\;\;\
Q_1(x)=x-\alpha_1,$$
\begin{equation}\label{op}
xQ_n(x)=Q_{n+1}(x)+\alpha_{n+1}Q_n(x)+\omega_nQ_{n-1}(x),
\end{equation}
for $n\geq 1$. If such a spectral distribution is unique (e.g., if
the uniform boundedness condition $(b')$ is fulfilled), the
spectral distribution $\mu$ is determined by the identity:
\begin{equation}\label{v3}
G_{\mu}(x)=\int_{R}\frac{\mu(dy)}{x-y}=\frac{1}{x-\alpha_1-\frac{\omega_1}{x-\alpha_2-\frac{\omega_2}
{x-\alpha_3-\frac{\omega_3}{x-\alpha_4-\cdots}}}}=\frac{Q_{n-1}^{(1)}(x)}{Q_{n}(x)}=\sum_{l=1}^{n}
\frac{A_l}{x-x_l},
\end{equation}
where $G_{\mu}(x)$ is called the Stieltjes transform and $A_l$ is
the coefficient in the Gauss quadrature formula corresponding to
the roots $x_l$ of polynomial $Q_{n}(x)$ and where polynomials
$\{Q_{n}^{(1)}\}$ are defined
recurrently as\\
        $Q_{0}^{(1)}(x)=1$,\\
    $Q_{1}^{(1)}(x)=x-\alpha_2$,\\
    $xQ_{n}^{(1)}(x)=Q_{n+1}^{(1)}(x)+\alpha_{n+2}Q_{n}^{(1)}(x)+\omega_{n+1}Q_{n-1}^{(1)}(x)$,

    for $n\geq 1$.

Now if $G_{\mu}(x)$ is known, then the spectral distribution
$\mu$ can be recovered from $G_{\mu}(x)$ by means of the
Stieltjes inversion formula:
\begin{equation}\label{m1}
\mu(y)-\mu(x)=-\frac{1}{\pi}\lim_{v\longrightarrow
0^+}\int_{x}^{y}Im\{G_{\mu}(u+iv)\}du.
\end{equation}
Substituting the right hand side of (\ref{v3}) in (\ref{m1}), the
spectral distribution can be determined in terms of $x_l,
l=1,2,...$, the roots of the polynomial $Q_n(x)$, and  Guass
quadrature constant $A_l, l=1,2,... $ as
\begin{equation}\label{m}
\mu=\sum_l A_l\delta(x-x_l)
\end{equation}
 ( for more details see Ref. \cite{obah,tsc,st,obh}.)

In the  following, we show the above mentioned isometry $U$ from
the orthogonal polynomial into interacting Fock space in infinite
graphs associated with Hermite, Laguerre and Tchebichef
polynomials, simply by replacing the raising (lowering) part of
adjacency matrix, i.e., $A^{+}$ ($A^{-}$) by the corresponding
first order differential operators
$B_{-}(n)$($A_{-}(n)$)introduced by one of the authors in Ref.
\cite{jf} with the following recursion relations
\begin{equation}
B_{-}(n)Q_{n-1}(x)=Q_{n}(x) ;\;\;\;\;\;\;\;\
A_{-}(n)Q_{n}(x)=E(n)Q_{n-1}(x),
\end{equation}
where  $E(n)=n$, for Hermite, $E(n)=n(n-\gamma), \gamma\
>-1$, for Laguerre and $E(n)=n^2$ for Tchebichef. Now by assuming
the canonical  isometry $UQ_{n}(x)=\sqrt{\beta_n}\ket{\phi_n}$
between the corresponding Hilbert spaces,
$A^+=UB_{-}(n)U^{\star}$, $A^-=UA_{-}(n)U^{\star}$ and the
corresponding operators  $\omega_n=E(n)$, we have\\
\\
 for Laguerre polynomial  $Q_{n}(x)=L_n(x) $
 $$
A^+=U(-x\frac{d}{dx}+x-n)U^{\star}  ;\;\;\;\;\
 A^-=U(x\frac{d}{dx}-n)U^{\star},
 $$
 for Hermite polynomial $Q_{n}(x)=H_n(x)$
$$
A^+=U(-\frac{d}{dx}+x)U^{\star}  ;\;\;\;\;\
 A^-=U(\frac{d}{dx})U^{\star}.
$$
for Tchebichef polynomial $Q_{n}(x)=T_n(x)$
$$
A^+=U(-(1-x^2)\frac{d}{dx}+nx)U^{\star}  ;\;\;\;\;\
 A^-=U((1-x^2)\frac{d}{dx}-nx)U^{\star}.
$$

Finally, using the quantum decomposition relations ( 3-7,8,9) and
the recursion relation (\ref{op}) of polynomial $Q_n(x)$, the
other matrix elements $\label{cw1} \braket{\phi_{k}}{A^m\mid
\phi_0}$ can be written as
\begin{equation}\label{cw1}
\braket{\phi_{k}}{A^m\mid
\phi_0}=\frac{1}{\sqrt{\omega_1\omega_2\cdots \omega_{k}
}}\int_{R}x^{m}Q_{k}(x)\mu(dx),  \;\;\;\;\ m=0,1,2,....
\end{equation}
Our main goal in this paper is the evaluation of amplitude
probability  for continuous-time quantum walk by using
Eq.(\ref{cw1}) such that we have
\begin{equation} \label{v4}
q_{k}(t)=\braket{\phi_{k}}{e^{-iAt}\mid
\phi_0}=\frac{1}{\sqrt{\omega_1\omega_2\cdots\omega_{k}}}\int_{R}e^{-ixt}Q_{k}(x)\mu(dx),
\end{equation}
where $q_{k}(t)$ is the amplitude of observing the particle at
level $k$ at time $t$. The conservation of probability
$\sum_{k=0}{\mid q_{k}(t)\mid}^2=1$ follows immediately from
Eq.(\ref{v4}) by using the completeness relation of orthogonal
polynomials $Q_n(x)$.
 Also by using Eq.(\ref{v4}) and
(\ref{v5},\ref{v6},\ref{v7}), we obtain the following recurrence
relation for the amplitude $p_{k}(t)$:
\begin{equation} \label{p}
i\frac{dq_{k}(t)}{dt}=\sqrt{\omega_{k+1}}q_{k+1}(t)+\alpha_{k+1}q_{k}(t)+
\sqrt{\omega_{k}}q_{k-1}(t).
\end{equation}
Obviously evaluation of $q_{k}(t)$ leads to the determination of
the amplitudes at sites belonging to the  stratum $V_k$, as it is
proved in the appendix I, the  walker has the same amplitude at
sites belonging to the same stratum, i.e., we have
$q_{ik}(t)=\frac{q_{k}(t)}{\mid V_k\mid}, i=0,1,...,\mid V_k\mid$,
where $q_{ik}(t)$ denotes the amplitude walker at $i$th site of
$k$th stratum.\\
Obviously for finite graphs, the formula (\ref{v4}) yields
\begin{equation}\label{fin}
q_{k}(t)=\frac{1}{\sqrt{\omega_1\omega_2\cdots\omega_{k}}}\sum_{l}A_le^{-ix_lt}Q_{k}(x_l),
\end{equation}
where by straightforward  calculation one can  evaluate  the
average probability for the finite graphs as
\begin{equation}
P(k)=\lim_{T\rightarrow \infty}\frac{1}{T}\int_{0}^{T}\mid
q_{k}(t)\mid
^2dt=\frac{1}{\omega_1\omega_2\cdots\omega_{k}}\sum_{l}A_l^2Q_{k}^2(x_l).
\end{equation}
 Finally using Gauss quadrature formula \cite{tsc}
\begin{equation}\label{ap}
\int f(x)\mu(dx)=\sum_{l=1}^{n}A_k f(x_l),
\end{equation}
where  the constants $A_l$ and $x_l$ are the same as those
appearing in formula(\ref{v3}), we can approximate an infinite QD
graph corresponding to the sequence of orthogonal polynomials $
\{Q_0, Q_1, Q_2,...\}$, with the finite $n$-stratum QD graph of
spectral distribution obtained from Stieltjes transform
(\ref{v3}) with constant $A_k$ and $x_k$, and vice versa.
Therfore, using Gauss quadrature formula, we can interchange
formulas (\ref{v4}) and (\ref{fin}) of finite and infinite graphs.
 This approximation can be very useful  in obtaining the
large time asymptotic behavior of quantum walk on graphs with
finitely many vertices.

Since, studying the large time behavior of quantum walks on
finite or infinite graphs by this method, naturally leads us to
consider the behavior of integrals of the form
 \begin{equation}\label{int1}
I(t)=\int_{a}^{b}e^{-it\phi(x)}g(x)dx   \;\;\;\;\;\ a\leq x\leq b
\end{equation}
as $t$ tends to infinity. There is a well-developed theory of the
asymptotic expansion of integrals known as method of stationary
phase approximation or WKB  \cite{cbs,nbrh}, which allows us to
determine, very precisely, the leading terms in the expansion of
the integral in terms of simple functions of $t$ (such as inverse
powers of $t$).  In  section $7$ we will study asymptotic
behavior of some finite and infinite graphs at distant time by
using Gauss quadrature formula and WKB approximation.
\section{ The direct product of QD finite
graphs} In this section, we study continuous-time quantum walk on
direct product of QD finite graphs. Let $G_i, i=1,2,..,n$ be
graphs of finite vertices with the corresponding adjacency
matrices $A_i, i=1,2,...,n$. Then their direct product
 \be\label{p1}\label{gp}
G_1\otimes\cdots \otimes G_n,
 \ee
is a graph with the following  adjacency matrix  $A$:
  \be\label{a1}
  A=\sum_{j=1}^{n}
  I\otimes\cdots\otimes A_j\otimes\cdots\otimes I
  \ee
where the $j-th$ term in the sum has $A_j$ appearing in the $j-th$
place in the tensor product.\\
In this case, for the amplitude of the ground state
$\ket{\phi_0}=\ket{\phi_{0}^{(1)}}\ket{\phi_{0}^{(2)}}\cdots\ket{\phi_{0}^
{(n)}}$, similar to  Eq.(\ref{v2}), we have
\begin{equation}
\braket{\phi_0}{A^m\mid
\phi_0}=\int\cdots\int(x_1+x_2+\cdots+x_n)^m
\mu(dx_1)\mu(dx_2)\cdots\mu(dx_n).
\end{equation}
Thus the  amplitude of the ground state can be written as
$$
q_0(t)=\braket{\phi_0}{e^{-itA}\mid \phi_0}=\int\cdots\int
e^{-itx_1}e^{-itx_2}\cdots e^{-itx_n}
\mu(dx_1)\mu(dx_2)\cdots\mu(dx_n)
$$
\begin{equation}\label{pr}
=\braket{\phi_{0}^{(1)}}{e^{-itA_1}\mid
\phi_{0}^{(1)}}\braket{\phi_{0}^{(2)}}{e^{-itA_2}\mid
\phi_{0}^{(2)}}\cdots \braket{\phi_{0}^{(n)}}{e^{-itA_n}\mid
\phi_{0}^{(n)}}.
\end{equation}
Therefore the amplitude of the walker at  ground state $q_0(t)$
($q_k(t)$ the amplitude of the walker at stratum $k$ and time $t$
) of product graph can be obtained simply by multiplying the
corresponding amplitudes of sub-graphs. Also comparing equations
(\ref{v4}) and (\ref{pr}) one can determine the spectral
distribution $\mu(dx)$ and Stieltjes transform
$G_{\mu}(x)$(consequently all amplitudes of product graph),
provided that the product graph possesses quantum distribution,
too. In the following, we will obtain the required conditions for
obtaining  QD graphs  from the product of two given QD graphs. As
we will see it is quite possible to obtain QD  graphs with the
same but non-isomorphic spectral distribution (having different
adjacency matrices, such that the adjacency matrix of product one
may not be tensor product of adjacency matrix of multiplicand
graphs)
from the products of QD graphs.\\
Considering two QD graphs, $G_1$ with
$\omega^{1}_1,\omega^{1}_2,...,\omega^{1}_n;
\alpha^{1}_2,\alpha^{1}_3,...,\alpha^{1}_{n+1}$ and $G_2$ with\\
$\omega^{2}_1,\omega^{2}_2,...\omega^{2}_m;
\alpha^{2}_2,\alpha^{2}_3,...,\alpha^{2}_{m+1} $, we obtain the
following QD graph from their product

$$
\ket{\phi_0}=\ket{\phi^{1}_0}\ket{\phi^{2}_0},
$$
$$
\ket{\phi_1}=\frac{1}{\sqrt{\omega^{1}_1+\omega^{2}_1}}
(\sqrt{\omega^{1}_1}\ket{\phi^{1}_1}\ket{\phi^{2}_0}+\sqrt{\omega^{2}_1}\ket{\phi^{1}_0}\ket{\phi^{2}_1}),
$$
$$
\ket{\phi_2}=\frac{1}{\sqrt{\omega^{1}_1\omega^{1}_2+4\omega^{1}_1\omega^{2}_2+\omega^{2}_1\omega^{2}_2}}
(\sqrt{\omega^{1}_1\omega^{1}_2}\ket{\phi^{1}_2}\ket{\phi^{2}_0}+2\sqrt{\omega^{1}_1\omega^{2}_1}
\ket{\phi^{1}_1}\ket{\phi^{2}_1}+\sqrt{\omega^{2}_1\omega^{2}_2}\ket{\phi^{1}_0}\ket{\phi^{2}_2}),
$$
$$
\vdots
$$
\begin{equation}
\ket{\phi_{mn-1}}=\ket{\phi^{1}_n}\ket{\phi^{2}_m},
\end{equation}
with
$$
\omega_1=\omega^{1}_1+\omega^{2}_1,  \;\;\;\;\
\omega_2=\frac{\omega^{1}_1\omega^{2}_2+4\omega^{1}_1\omega^{2}_1+\omega^{2}_1\omega^{2}_2}
{\omega^{1}_1+\omega^{2}_1}, \;\;\;\;\;\
$$
$$
\omega_3=\frac{\omega^{1}_1\omega^{1}_2\omega^{1}_3+
9(\omega^{1}_1\omega^{1}_2\omega^{2}_1+\omega^{1}_1\omega^{2}_1\omega^{2}_2)+\omega^{2}_1\omega^{2}_2\omega^{2}_3}
{\omega^{1}_1\omega^{2}_2+4\omega^{1}_1\omega^{2}_1+\omega^{2}_1\omega^{2}_2},
$$
$$
\omega_4=\frac{\omega^{1}_1\omega^{1}_2\omega^{1}_3\omega^{1}_4+16\omega^{1}_1\omega^{1}_2\omega^{1}_3\omega^{2}_1+
36\omega^{1}_1\omega^{1}_2\omega^{2}_1\omega^{2}_2+16\omega^{1}_1\omega^{2}_1\omega^{2}_2\omega^{2}_3+
\omega^{2}_1\omega^{2}_2\omega^{2}_3\omega^{2}_4}
{\omega^{1}_1\omega^{1}_2\omega^{1}_3+
9(\omega^{1}_1\omega^{1}_2\omega^{2}_1+\omega^{1}_1\omega^{2}_1\omega^{2}_2)+\omega^{2}_1\omega^{2}_2\omega^{2}_3},...
$$
$$
 \alpha_2= \alpha^1_2= \alpha^2_2, \;\;\;\;\
\alpha_3=\alpha^1_2+ \alpha^2_2=\alpha^1_3=\alpha^2_3, \;\;\;\;\
$$
\begin{equation}
\alpha_4=\alpha^1_2+ \alpha^2_3=\alpha^1_3+
\alpha^2_2=\alpha^1_4=\alpha^2_4,...,
\alpha_{k+1}=k\alpha^1_2.....
\end{equation}
 Now, by acting operators $A^{+}, A^{-}$ and $A_{0}$ on the quantum state of the
 product graph (\ref{gp}), we get the following required conditions between the  QD
parameters of  multiplicant graphs in order to obtain a QD graph
from their Cartesian product
$$
\omega^{1}_2+2\omega^{2}_1=2\omega^{1}_1+\omega^{2}_2, \;\;\;\;\
2(\omega^{1}_3+3\omega^{2}_1)=3(\omega^{1}_2+\omega^{2}_2),
$$
$$
\omega^{1}_3+3\omega^{2}_1=3\omega^{1}_1+\omega^{2}_3,  \;\;\;\;\
\omega^{1}_4+4\omega^{2}_1=4\omega^{1}_1+\omega^{2}_4,
$$
\begin{equation}
3\omega^{1}_4+12\omega^{2}_1=4\omega^{1}_3+6\omega^{2}_2,
\;\;\;\;\
6\omega^{1}_2+4\omega^{2}_3=12\omega^{1}_1+3\omega^{2}_4,
\end{equation}
$$
\vdots
$$
Obviously it is difficult to solve the above equations in general.
But one can show that these equations hold for graphs which are
themselves  products of some elementary QD graphs with only one,
two  or at most three nonzero QD parameters.\\

Thus, in the following,  we will consider two important classes of
direct product, namely  class $A$ and class $B$ with some
relevant examples  for each class.

 \textbf{Class A.}
Class A graphs consist of product of an arbitrary numbers of QD
graphs with only two  non-vanishing QD
 parameters
 \begin{equation}
  \omega_1=a ,\;\;\;\; \omega_2=\omega_3=\cdots=0 ; \;\;\;\;\;\
   \alpha_{1}=0, \;\;\;\;\;\  \alpha_{2}=b, \;\;\;\;\;\ \alpha_{3}=\alpha_{4}=\cdots=0,
\end{equation}
where  for their $n$-fold direct product we have
$$
\ket{\psi_0}=\ket{\phi_0}\ket{\phi_0}\cdots\ket{\phi_0}=\ket{\phi_0}^{\otimes
n }
$$
$$
\ket{\psi_1}=\frac{1}{\sqrt{C_{1}^{n}}}\sum_{permutation}\ket{\phi_0}\ket{\phi_0}
\cdots\ket{\phi_0}\ket{\phi_1},
$$
$$
\ket{\psi_2}=\frac{1}{\sqrt{C_{2}^{n}}}\sum_{permutation}\ket{\phi_0}\ket{\phi_0}
\cdots\ket{\phi_0}\ket{\phi_1}\ket{\phi_1},
$$
$$
\vdots
$$
\begin{equation}
\ket{\psi_n}=\ket{\phi_1}\ket{\phi_1}\cdots\ket{\phi_1}=\ket{\phi_1}^{\otimes
n }.
\end{equation}
Application of operators $A^{+}, A^{-}$ and $A_{0}$ on quantum
states of the product graph given above yields for its QD
parameters:
 $\omega_k'=k(n-(k-1))a$, for
$k=1,2,3,...,n$, and $\alpha'_{k}=(k-1)b$,\\
where the QD property of the product graph imposes the following
conditions: $\mid V_k\mid-1\geq \alpha'_{k+1}$ and $\frac{\mid
V_k\mid \alpha'_{k+1}}{2}=d$, where $d$ is an integer.\\ Here we
give two non-isomorphic graphs out of many prototypes of class A:
 \\
 $
graph1.\mid V_k\mid=\omega'_1\omega'_2...,\omega'_k,(tree)\\
graph2.\mid V_1\mid=na, ;\;\;\ \mid V_2\mid=2n^2, ;\;\;\;\ \mid
 V_3\mid=3!an(n-2),...
 $.

\textbf{$n$-cube as an example of class A}\\
 As an example, we obtain $n$-cube as the direct product of $K_2$
 graphs by choosing  $a=1$, and $b=0$. Hence as we will see  in the following
 that its amplitude  $\ket{\psi_{t}}$ can be written as the product of
 the $K_2$ ones.  Since its  adjacency matrix can be written
 as the Cartesian  product of the $K_2$ graphs therefore
 \be
A=\frac{1}{n}\sum_{i=1}^{n} I_{2}\otimes\cdots\otimes
\sigma_{x}\otimes\cdots\otimes I_{2} \ee where the $i$th term in
the sum has $\sigma_{x}$ (Pauli matrix) appearing in the $i$th
place in the tensor product. Therefore, we have
$$
 exp(-iAt)=\prod_{i=1}^{n} I_{2}\otimes\cdots\otimes
e^{-it\sigma_{x}/n}\otimes\cdots\otimes I_{2}$$ \be
=e^{-it\sigma_{x}/n}\otimes\cdots\otimes
e^{-it\sigma_{x}/n}=(e^{-it\sigma_{x}/n})^{\otimes n} \ee where
$B^{\otimes n}$ is the tensor product of $n$ copies of $B$. On
the other hand  for $K_2$ we have
\begin{equation}
\omega_1=1, \;\;\;\;\ \omega_2=\omega_3=\cdots=0, \;\;\;\;\
\alpha_1=\alpha_2=\cdots=0
\end{equation}
and
\begin{equation}
G_(x)=\frac{x}{x^2-1}, \;\;\;\;\;\
\mu=\frac{1}{2}(\delta(x-1)+\delta(x+1)).
\end{equation}
Hence, for the  amplitude of observing the walker  at 0-th
stratum of $n$-cube  at  time $t$, we have
$$
q_0(t)=\braket{\phi_0}{e^{-itA}\mid
\phi_0}=\braket{\phi_{0}^{(1)}}{e^{-it\sigma_x/n}\mid
\phi_{0}^{(1)}}\cdots
\braket{\phi_{0}^{(n)}}{e^{-it\sigma_x/n}\mid \phi_{0}^{(n)}}
$$
\begin{equation}
=\cos{(t/n)}\cdots \cos{(t/n)}=\cos^{n}{(t/n)}
\end{equation}
and  the other amplitudes can be calculated by the prescription
explained above.

\textbf{Class  B.} Class B graphs consist of product of an
arbitrary numbers of QD graphs with only four non-vanishing QD
 parameters
 \begin{equation}
  \omega_1=\omega_2=a, \;\;\;\; \omega_3=\cdots=0;
  \;\;\;\;\;\\
   \alpha_{1}=0, \;\;\;\ \alpha_{2}=b, \;\;\;\ \alpha_{3}=2b, \;\;\;\ \alpha_{4}=\alpha_{5}=\cdots=0,
\end{equation}
where  for their $n$-fold direct product we have
$$
\ket{\psi_0}=\ket{\phi_0}\ket{\phi_0}\cdots\ket{\phi_0}=\ket{\phi_0}^{\otimes
n }
$$
$$
\ket{\psi_1}=\frac{1}{\sqrt{C_{1}^{n}}}\sum_{permutation}\ket{\phi_0}\ket{\phi_0}
\cdots\ket{\phi_0}\ket{\phi_0}\ket{\phi_1},
$$
$$
\ket{\psi_2}=\frac{1}{\sqrt{C_{1}^{n}+(2!)^2C_{2}^{n}}}(2!\sum_{permutation}\ket{\phi_0}\ket{\phi_0}
\cdots\ket{\phi_1}\ket{\phi_1})+\sum_{permutation}\ket{\phi_0}\ket{\phi_0}
\cdots\ket{\phi_0}\ket{\phi_2}),
$$
$$
\vdots
$$
\begin{equation}
\ket{\psi_{2n}}=\ket{\phi_2}\ket{\phi_2}\cdots\ket{\phi_2}=\ket{\phi_2}^{\otimes
n }.
\end{equation}
Similarly, application of the operators $A^{+}, A^{-}$ and $A_{0}$
on the quantum states of the product graph given above yields for
its QD parameters: $\omega_{2k+1}'=(2k+1)(n-k)a$, for
$k=0,1,2,3,...,n$, $\omega_{2k}'=k(2n-(2k-1))a$,for
$k=1,2,3,...,n$ and $\alpha'_{n+1}=nb$ together with the
constraint similar to the class A ones plus an  extra constraint
$\omega_1=\omega_2$ coming from its QD property.  In the
following, out of many
prototypes of class B, we give the angular momentum graphs as an example. \\
\textbf{Example. Angular momentum graphs}\\
Angular momentum graphs  can be obtained from symmetric tensor
product of QD graphs of class A with  $a=2$ and $b=1$, called
vector graphs ( as the angular momentum l  can be obtained from
 symmetric tensor product of vectors). Again the amplitude
$\ket{\psi_{t}}$ can be  written as the product of vector graphs,
since its  adjacency matrix can be written as Cartesian  product
of  vector graphs

\begin{equation} A=\frac{1}{n}\sum_{i=1}^{n}
I_{7}\otimes\cdots\otimes A'\otimes\cdots\otimes I_{7}
\end{equation}
where the $i$th term in the sum has $A'$ (adjacency matrix of
vector graphs ) appearing in the $i$-th place in the tensor
product. On the other hand,  for vector graph we have
\begin{equation}
\omega_1=\omega_2=2, \;\;\;\ \omega_3=\omega_4=\cdots=0; \;\;\;\;\
\alpha_1=0, \;\;\;\;\ \alpha_2=1, \;\;\;\;\ \alpha_3=2, \;\;\;\;\
\alpha_4=\alpha_4=\cdots=0
\end{equation}
and
\begin{equation}
G_(x)=\frac{x^2-3x}{x^3-3x^2-2x+4}, \;\;\;\;\;\
\mu=\frac{2}{5}\delta(x-1)+\frac{3-\sqrt{5}}{10}\delta(x-(1+\sqrt{5}))
+\frac{3+\sqrt{5}}{10}\delta(x-(1-\sqrt{5})).
\end{equation}
Hence, for $l=n$ angular momentum graph with the following QD
parameters

\begin{equation}
\omega_1=2n, \;\;\;\
\omega_1=2(2n-1),\cdots,\omega_k=k(2n-k+1),\cdots; \;\;\;\
\alpha_{k}=k-1,
\end{equation}
the  amplitude of observing the walker  at 0-th stratum of at time
$t$ becomes
$$ q_0(t)=\braket{\phi_0}{e^{-itA}\mid
\phi_0}=\braket{\phi_{0}^{(1)}}{e^{-itA'/n}\mid
\phi_{0}^{(1)}}\cdots \braket{\phi_{0}^{(n)}}{e^{-itA'/n}\mid
\phi_{0}^{(n)}}
$$
$$
=\frac{1}{5}e^{-it/n}(2+3\cos(\sqrt{5}t/n)+i\sqrt{5}\sin(\sqrt{5}t/n))\cdots
\frac{1}{5}e^{-it/n}(2+3\cos(\sqrt{5}t/n)+i\sqrt{5}\sin(\sqrt{5}t/n))
$$
\begin{equation}
=(\frac{1}{5}e^{-it/n}(2+3\cos(\sqrt{5}t/n)+i\sqrt{5}\sin(\sqrt{5}t/n)))^n,
\end{equation}
and  the other amplitudes can be calculated by the prescription
explained above.
\section{Some examples of QD graphs}
In this section we provide some examples of finite and infinite QD
graphs and using the spectral distribution we calculate the
relevant amplitudes of continuous time quantum walks on theses
graphs. As far as the authors know, theses examples exhaust all
known graphs which have been studied by researcher in quantum walk
until now. First we begin with finite QD graphs:
 \subsection{Complete graph $K_n$}
A complete graph with n vertices (denoted Kn) is a graph with n
vertices in which each vertex is connected to the others (with
one edge between each pair of vertices). Therefore, it is a
two-stratum QD graph with one non-vanishing QD parameter
$\omega_1=n-1 $ and spectral distribution
\begin{equation}
G_{\mu}(x)=\frac{x-n+2}{x^2-(n-2)x-n+1}, \;\;\;\
\mu=\frac{n-1}{n}\delta(x+1)+\frac{1}{n}\delta(x-(n-1)),
\end{equation}
that yields the following amplitudes at time t of the quantum
walker

$$
q_0(t)=\braket{\phi_0}{e^{-iAt/(n-1)}\mid\phi_0}=\int_{R}e^{-ixt/(n-1)}\mu(dx)
$$
\begin{equation}
=\frac{n-1}{n}e^{\frac{it}{n-1}}+\frac{1}{n}e^{-it}=\frac{1}{n}(e^{-it}+(n-1)e^{\frac{it}{n-1}})
\end{equation}
 $$
q_1(t)=\braket{\phi_1}{e^{-iAt/(n-1)}\mid\phi_0}=\int_{R}\frac{x}{n-1}e^{-ixt/(n-1)}\mu(dx)
$$
\begin{equation}
=\frac{\sqrt{n-1}}{n}(e^{-it}-e^{it/(n-1)}),
\end{equation}
where the results thus obtained are in agreement with those of
Ref.\cite{abtw}.
\subsection{Cycle graph $C_n$}
A cycle graph or cycle is a graph that consists of some number of
vertices connected in a closed chain. The cycle graph with n
vertices is denoted by $ C_n$, and quantum walk on them turns out
to be different for odd and even $n$, hence we treat them
separately.\\
 \textbf{Odd $n$.} For odd $n=2m+1$, $C_{2m+1}$ graph  consists of
 $m+1$ starata, and its QD parameters are
 \begin{equation}
  \omega_1=2 ,\;\;\;\; \omega_2=\omega_3=\cdots\omega_{m}=1 ; \;\;\;\;\;\
  \alpha_1=\alpha_2=\cdots=0, \;\;\ \alpha_{m+1}=1.
\end{equation}
It is straightforward to show that it has the following spectral
distribution
\begin{equation}\label{d1}
\mu=\frac{1}{2m+1}\delta(x-2)+\frac{2}{2m+1}\sum_{l=1}^{2m}\delta(x-2\cos(\frac{2l\pi}{2m+1})).
\end{equation}
Therefore, the amplitude probability for walker at time $t$ and
$0$th stratum is
$$
q_0(t)=\braket{\phi_0}{e^{-iAt/2}\mid\phi_0}=\int_{R}e^{-ixt/2}\mu(dx)
$$
\begin{equation}
=\frac{1}{2m+1}e^{-it}+\frac{2}{2m+1}\sum_{l=1}^{m}e^{-it\cos{2l\pi/(2m+1)}}.
\end{equation}
Using Eq.(\ref{v4}), one can calculate the other amplitudes,
where the results obtained are in agreement with those of
Ref.\cite{abtw}.\\
 \textbf{Even $n$.} For even $n=2m$, $C_{2m}$ graph  consists of
 $m+1$ starata, and its QD parameters are
 \begin{equation}
  \omega_1=2 ,\;\;\;\; \omega_2=\omega_3=\cdots\omega_{m-1}=1 ; \;\;\
  \omega_{m}=2;   \;\;\;\;\;\
  \alpha_1=\alpha_2=\cdots=0.
\end{equation}
One can  straightforwardly show that it has the following
spectral distribution
\begin{equation}\label{d2}
\mu=\frac{1}{2m}(\delta(x-2)+\delta(x+2))+\frac{2}{2m}\sum_{l=1,
l\neq m }^{2m-1}\delta(x-2\cos(\frac{2l\pi}{2m})).
\end{equation}
Similarly, the amplitude probability for walker at time $t$ and
$0-th$ stratum is
$$
q_0(t)=\braket{\phi_0}{e^{-iAt/2}\mid\phi_0}=\int_{R}e^{-ixt/2}\mu(dx)
$$
\begin{equation}
=\frac{1}{m}\cos{t}+\frac{1}{m}\sum_{l=1}^{m}e^{-it\cos{2l\pi/(2m)}}.
\end{equation}
Again one can calculate the other amplitudes by using
Eq.(\ref{v4}), where the results are in agreement with those of
Ref. \cite{abtw}.\\

 In the limit of large $n$, the roots $x_l=2\cos(\frac{2\pi l}{2m})$
  reduce to $x_l=2\cos(\pi x)$ with $x=\lim_{m,l\rightarrow\infty}\frac{l}{m}$
and both spectral distributions given in (\ref{d1}) and
(\ref{d2}), irrespective of whether $n$ is even or odd, reduce to
 continuous spectral distribution  $\mu(x)dx$ with $\mu(x)$
$$
 \mu(x)=\frac{1}{\pi}\int_{0}^{\pi}
dy\delta(x-2\cos(y))
$$
 $$=\frac{1}{\pi}\int_{0}^{\pi}
dy\frac{\delta(y-\arccos(x/2))}{2\sin(y)}
$$
$$
=\frac{1}{\pi}\int_{0}^{\pi}
dy\frac{\delta(y-\arccos(x/2))}{2\sin(y)}
$$
\begin{equation}
=\frac{1}{\pi}\frac{1}{\sqrt{4-x^2}},  \;\;\;\;\;\;\ -2\leq x\leq
2,
\end{equation}
which is the same as the continuous  spectral distribution of
infinite line graph $\textsc{Z}$.
\subsection{ Tchebichef graphs}
By choosing Tchebichef polynomials of first kind (second kind
)with scaling factor $\frac{1}{2^m}$ as orthogonal polynomials
appearing in recurrence relation (\ref{op}), i.e.,
$Q_n(x)=2^{(m-1)n+1}T_{n}(x/2^m)$ $(2^{(m-1)n} U_n(x/2^m))$, one
can obtain a class of finite and infinite QD graphs of Tchebichef
type, with QD parameters $\omega_1=2^{2(m-1)+1}, \;\;\
\omega_k=2^{2(m-1)}, \;\ k=2,3,...$, and $\alpha_k=0, \;\
k=1,2,3,...$ $(\omega_k=2^{2(m-1)}, \;\ k=1,2,... ; \;\;\
\alpha_k=0, \;\ k=1,2,... )$, such that its Stieltjes transform
of spectral distribution becomes
$G_\mu(x)=\frac{1}{n}\frac{T'_n(\frac{x}{2^m})}{T_{n}(\frac{x}{2^m})}$
$(\frac{1}{2^{m-1}}\frac{U_n(\frac{x}{2^m})}{U_{n+1}(\frac{x}{2^m})})$.

Therefore its spectral distribution can be written as
$\mu=\frac{1}{n}\sum_{l=0}\delta(x-2^m\cos{\frac{(2l+1)\pi}{2n}})$
$(\frac{1}{n+2}\sum_{l=0}\sin^2(\frac{l\pi}{n+2})\delta(x-2^m\cos{\frac{l\pi}{n+2}}))$,
where $2^m\cos{\frac{(2l+1)\pi}{2n}}, l=1,2,... $
$(2^m\cos(\frac{l\pi}{n+2}),l=1,2,\cdots  )$ are  roots of the
Tchebishef polynomial, $T_{n}(x/2^m)$ $(U_{n+1}(x/2^m))$. Now,
using Eq.(\ref{v4}) the amplitude of observing walker at
different stratum at time $t$ can be calculated as
$$
q_{0}(t)=\frac{1}{n}\sum_{l=0}e^{-i2^mt\cos(\frac{(2l+1)\pi}{2n})},
$$
\begin{equation}\label{tc1}
q_{k}(t)=\frac{\sqrt{2}}{n}\sum_{l=0}e^{-i2^mt\cos(\frac{(2l+1)\pi}{2n})}
\cos(\frac{k(2l+1)\pi}{2n}), \;\;\;\ k\geq 1,
\end{equation}
for the first kind and
$$
q_{0}(t)=\frac{2}{n+2}\sum_{l=0}\sin^2(\frac{l\pi}{n+2})e^{-i2^mt\cos(\frac{l\pi}{n+2})},
$$
\begin{equation}\label{tc2}
q_{k}(t)=\frac{2}{n+2}\sum_{l=0}\sin^2(\frac{l\pi}{n+2})\cos(\frac{(k+1)l\pi}{n+2})
e^{-i2^mt\cos(\frac{l\pi}{n+2})} , \;\;\;\ k\geq 1,
\end{equation}
for the second kind.

Finally, in the limit of large $n$,  the absolutely continuous
part of  spectral distribution $\mu(x)$ reads as
$\mu(x)=\frac{1}{\pi}\frac{1}{\sqrt{2^{2m}-x^2}}$
$(\frac{1}{2\pi}\sqrt{2^{2m}-x^2})$ such that, $-2^m\leq x\leq
2^m$(its derivation is similar to that of cycle graph $C_n$) and
the amplitude $p_k(t)$ given in (\ref{tc1}) and (\ref{tc2})
reduce to
$$
q_{0}(t)=J_{0}(2^{m-1}t),
$$
$$
q_{k}(t)=\frac{1}{\sqrt{\omega_1\omega_2\cdots\omega_k}}\int_{-2^m}^{2^m}2^{k(m-1)+1}T_k(x/2^m)e^{-ixt/2}\mu(dx)
$$
$$
=\frac{2}{\sqrt{2}}[\frac{1}{2\pi}\int_{-\pi}^{\pi}\cos(k\psi)e^{-i2^{m-1}t\cos(\psi)}d\psi]
$$
\begin{equation}\label{tc3}
=\sqrt{2}i^kJ_{k}(2^{m-1}t), \;\;\;\ k\geq 1,
\end{equation}
for the first kind and
$$
q_{k}(t)=\frac{1}{\sqrt{\omega_1\omega_2\cdots\omega_k}}\int
2^{(m-1)k} U_k(x/2^m)e^{-ixt/2}\mu(dx)
$$
$$
=\frac{1}{2\pi}\int_{-2^m}^{2^m}\frac{\sin((k+1)\cos^{-1}(x/2^m))}{\sqrt{1-(x/2^M)^2}}e^{-ixt/2}{\sqrt{2^{2m}-x^2}}dx
$$
$$
=\frac{2}{\pi}\int_{0}^{\pi}e^{-i2^mt\cos(\psi)}\sin((k+1)\psi)\sin(\psi)d\psi
$$
\begin{equation}\label{tc4}
=i^k(J_k(2^mt)+J_{k+2}(2^mt)), \;\;\;\ k=0,1,2,...,
\end{equation}
for the second kind, where $J_k(t)$ are the Bessel polynomials. In
deriving of (\ref{tc3}) and (\ref{tc4}), we have used the integral
representation of Bessel function after making the change of
variable $x=2^m\cos\psi$. Also, one can calculate the amplitudes
of infinite Tchebishef graphs  by using generating function of
Tchebishef and formula Eq.(\ref{v4}), where the obtained results
are in agreement with those given in (\ref{tc3}) and (\ref{tc4}).
As we will see in the following, some of the known finite and
infinite graphs can be obtained from Tchebishef graphs by
appropriate
choice of $m$ and the polynomials as:\\
\textbf{A.} For $m=1$ and Tchebishef polynomial of the second
kind we obtain finite path graph $\textsf{P}_n=\{0,1,2,... \}$,
where it is a $n$- vertex graph with $n-1$ edges  all on a single
open path. Its QD parameters and the amplitude of observing
walker at different stratum at time $t$  are
 $$
  \omega_1=\omega_2=\omega_3=\cdots=1 ; \;\;\;\;\;\
  \alpha_n=0
  $$
$$
q_0(t)=\frac{2}{n+2}\sum_{l=1}\sin^{2}(\frac{l\pi}{n+2})e^{-it\cos(\frac{l\pi}{n+2})},
$$
\begin{equation}
q_k(t)=\frac{2}{n+2}\sum_{l=1}^{n+1}\sin(l\pi/(n+2))\sin(\frac{k+1}{n+2}l\pi)e^{-it\cos(l\pi/(n+2))},
\;\;\ k\geq 1,
\end{equation}
 where the results thus obtained are in agreement with those of
 Ref.\cite{abt}.
Also, in the limit of large $n$,  $q_k(t)$  reduces to
\begin{equation}\label{pa1}
q_k(t)=i^k(J_k(t)+J_{k+2}(t)), \;\;\;\;\;\ k=0,1,....
\end{equation}

\textbf{B.} For $m=\frac{3}{2}$ and Tchebishef polynomial of the
second kind we obtain a sequence of graphs $G_n$ (for more details
see Ref.\cite{cfg}). The number of vertices in $G_n$ is
$2^{n+1}+2^n-2$. In Fig.1 we show $G_4$. In general, $G_n$
consists of two balanced binary trees of depth $n$ with the
$2^n$, $n$th-level vertices of the two trees pairwise identified.
For the quantum walk on $G_n$, one starts the waker at the root of
a tree and wants to calculate the probability of the presence of
the walker at the other vertices as function of time. One can
show that $G_n$ is QD graph with $(2n+1)$ strata, where stratum
$j$ consists of $2^j$ vertices for $j=1,2,...,n+1$ and
$2^{(2n+1-j)}$ for $j=n+1,...,2n+1$. Therefore, its QD parameters
and the amplitude of observing walker at different stratum at
time $t$ are
$$
  \omega_1=\omega_2=\omega_3=\cdots=2 ; \;\;\;\;\;\
  \alpha_1=\alpha_2=\cdots=0,
$$
$$
q_0(t)=\frac{2}{n+2}\sum_{l}\sin^{2}(\frac{l\pi}{n+2})
e^{-i2\sqrt{2}t\cos(\frac{l\pi}{n+2})},
$$
\begin{equation}
q_k(t)=\frac{2}{n+2}\sum_{l=1}^{n+1}\sin(l\pi/(n+2))
\sin(\frac{k+1}{n+2}l\pi)e^{-i2\sqrt{2}t\cos(l\pi/(n+2))}, \;\;\;\
k\geq 1.
\end{equation}
In the limit of large $n$, $p_k(t)$  reduce to
\begin{equation}\label{pa1}
q_k(t)=i^k(J_k(2\sqrt{2}t)+J_{k+2}(2\sqrt{2}t)), \;\;\;\;\;\
k=0,1,....
\end{equation}.
\textbf{C.} For $m=1$ and Tchebishef polynomial of the first kind
we obtain infinite line graphs $\textsc{Z}$ where its QD
parameters and  the amplitude of observing walker at different
stratum at time $t$ are
$$
 \omega_1=2 ,\;\;\;\; \omega_2=\omega_3=\cdots=1 ; \;\;\;\;\;\
  \alpha_k=0,
$$
$$
q_0(t)=J_0(t),
$$
\begin{equation}\label{tlp}
q_k(t)=\sqrt{2}i^{k}J_k(t), \;\;\;\;\  k\geq 1.
\end{equation}
The results thus obtained are in agreement with those of
Ref.\cite{koo}.\\
\subsection{The finite Hermit graph}
 Let $G$ consist of $n$ vertices such that its QD parameters,
 cardinality of stratums and Stieltjes transform are
$$
  \omega_1=n , \;\;\;\ \omega_2=n-1 , \;\;\;\ \omega_3=n-2 , \cdots, \omega_n=1 ,       \;\;\;\;\;\
  \alpha_1=\alpha_2=\cdots=0,
  $$
 $$
  \mid V_1\mid=n, \;\;\;\  \mid V_2\mid=n(n-1),  \cdots,    \mid
  V_n\mid=n!,
$$
\begin{equation}
G_\mu(x)=\frac{H_n(x)}{H_{n+1}(x)},
\end{equation}
where, $H_n(x)$ is Hermite polynomials. Then it will have the
following spectral distribution
\begin{equation}
\mu=\frac{1}{n+1}\sum_{l}\delta(x-x_l),
\end{equation}
where $x_l,\ l=1,2,...$ are the roots of  Hermite polynomial
$H_n(x)$. Again, using Eq.(\ref{v4}) or the recurrence relation
(\ref{p}) one can calculate all amplitudes, where in the
following we just quote the results for $n=3$. Its QD parameters
and spectral distribution are
$$
  \omega_1=3 ,\;\;\;\; \omega_2=2, \;\;\;\;\ \omega_3=1 ; \;\;\;\;\;\
   \alpha_1=\alpha_2=\cdots=0,
$$
\begin{equation}
\mu=\frac{1}{4}(\delta_{\sqrt{3+\sqrt{6}}}+
\delta_{\sqrt{3-\sqrt{6}}}+\delta_{-\sqrt{3+\sqrt{6}}}+\delta_{-\sqrt{3-\sqrt{6}}}),
\end{equation}
and the amplitude as
$$
q_o(t)=\frac{1}{2}(\cos(\sqrt{3+\sqrt{6}})t+\cos(\sqrt{3-\sqrt{6}})t),
$$
$$
q_1(t)=\frac{-i}{6}((\sqrt{3+\sqrt{6}})\sin(\sqrt{3+\sqrt{6}})t+
(\sqrt{3-\sqrt{6}})\sin(\sqrt{3-\sqrt{6}})t),
$$
$$
q_2(t)=\frac{1}{2\sqrt{6}}(\cos(\sqrt{3+\sqrt{6}})t-\cos(\sqrt{3-\sqrt{6}})t),
$$
\begin{equation}
q_3(t)=\frac{-i}{6\sqrt{2}}(\sqrt{3+\sqrt{6}}(\sqrt{3}-\sqrt{2})\sin(\sqrt{3+\sqrt{6}})t+
\sqrt{3-\sqrt{6}}(\sqrt{3}+\sqrt{2})\sin(\sqrt{3-\sqrt{6}})t).
\end{equation}
\subsection{The infinite Hermite graph}
By choosing  Hermite polynomial $H_n(x)$, as orthogonal
polynomial appearing in recurrence relation (\ref{op}), one can
obtain a class of infinite QD graphs of Hermite type with QD
parameters  $\omega_k=k$, and $\alpha_k=0$. Here we give two
non-isomorphic graphs out of many prototypes of  infinite Hermite
graph:\\
 $
Graph1. \mid V_k\mid=\omega_1\omega_2\cdots \omega_k=k!(tree) i.e., \kappa_{-}(y)=1 \\
 Graph2. \mid V_1\mid=1, \mid V_2\mid=2(\kappa_{-}(y)=1), \mid V_3\mid=3!(\kappa_{-}(y)=1), \mid
V_4\mid=3!(\kappa_{-}(y)=2),\mid V_5\mid=30(\kappa_{-}(y)=1), \mid
V_6\mid=6\times 30(\kappa_{-}(y)=1),... $.

 Here the amplitude $q_k(t)$  can be calculated rather trivially by using generating function of Hermite
 polynomials as follows:
$$
\sum_{k=0}\sqrt{\frac{1}{2k!\pi}}q_k(t)y^k=\int_{R} e^{-ixt}
\sum_{k}\frac{y^k}{k!} H_k(x)\mu(dx)=\int_{-\infty}^{\infty}
e^{-ixt} e^{xy-y^2/2}e^{-x^2/2}dx
$$
\begin{equation}\label{ge}
=e^{-y^2/2}\int_{-\infty}^{\infty}e^{-1/2(x^2+2x(it-y))}dx=e^{-\frac{1}{2}t(t+2iy)}.
\end{equation}
 Now, by comparing the terms of the expansion appearing on the left
 hand side of (\ref{ge}) with those of expansion on the right hand
 side we obtain
\begin{equation}\label{hp}
q_k(t)=\frac{(-i)^k t^k}{\sqrt{k!}} e^{-t^2/2},
\end{equation}
where the result thus obtained is exactly the same as  that of
WKB approximation, i.e, the method of stationary phase
 gives the exact result in infinite Hermite case.
Then, the probability distribution $\mid
q_k(t)\mid^2=\frac{(t^2)^k}{k!} e^{-t^2}$ is commonly called the
Poisson distribution.
 \subsection{The infinite Laguerre graph}
We take the  $Q_n(x)$ as a Laguerre polynomial, such that they
satisfy  the orthogonal polynomials appearing in the recurrence
relation (\ref{op}), i.e, $Q_n(x)=n!(-1)^n a^n
L_{n}^{(\gamma)}(\frac{x-b}{a})$. Then, one can obtain a class of
infinite QD graphs of Laguerre type , with QD parameters
$\omega_k=a^2k(k+\gamma)$, and choose $b=-a(1+\gamma)$,
$\alpha_1=0$ and $\alpha_{k+1}=2ka$ for $k\geq 1 $. Here we give
two non-isomorphic graphs out of many prototypes of  infinite
Hermite graph:\\
$
Graph1.\mid V_k\mid=\omega_1\omega_2\cdots \omega_k (tree) \;\;\ i.e., \kappa_{-}(y)=1\\
Graph2.\mid V_k\mid=k!a^2(1+\gamma)(2+\gamma)\cdots (k+\gamma),
\;\;\ (\kappa_{-}(y)=a). $

 By using Eq.(\ref{v4}), we obtain generating
functions for amplitude probability $q_k(t)$
$$
\sum_{k=0}\sqrt{\frac{(1+\gamma)(2+\gamma)\cdots
(k+\gamma)}{k!}}q_k(t)y^k=\int_{R} e^{-ixt} \sum_{k}(-1)^ka^k k!
L^{\gamma}_k(\frac{x-b}{a})\frac{y^k}{k!}\mu(dx)
$$
\begin{equation}
=\frac{1}{a}\int_{0}^{\infty}e^{-ixt}\frac{-\frac{y(x-b)}{a(1-y)}}{(1-y)^{\gamma+1}}
(\frac{x-b}{a})^{\gamma}e^{-(\frac{x-b}{a})}dx
=\frac{\gamma!e^{-ibt}}{(1+ia(1-y)t)^{\gamma+1}}.
\end{equation}
It is straightforward to show that $p_k(t)$ has the following
 form
\begin{equation}\label{lp}
q_k(t)=\sqrt{\frac{(\gamma+k)!}{k!\gamma!}}\frac{e^{-ibt}(iat)^k}{(1+iat)^{k+\gamma+1}}.
\end{equation}
Formula (\ref{lp}) implies that for distant time the amplitude
becomes proportional to the inverse of time $t$, i.e, we have
$q_{k}(t)\approx
\sqrt{\frac{(\gamma+k)!}{k!\gamma!}}\frac{e^{-ibt}}{t^{\gamma+1}}
$, which is the same as the one  obtained by  WKB approximation.

 \section{Asymptotic behavior of continuous-time quantum walk in the large time limit}
\subsection{Average moments of number of visiting strata}
 Studying the large time behavior of moments and variance as
function of $t$ for classical random walk and quantum walk is
important, since common properties for describing the behavior of
a walker are the exponents $\nu$, characterizing the scaling with
time t of the mean square end-to-end distance $\langle
R^2\rangle=\sqrt{\langle k^2\rangle-\langle k\rangle^2} \approx
t^{\nu}$(the exponent $\nu$ determine universality class). The
walk on infinite line, $\mid q_{k}(t)\mid^2$ has the form of a
binomial distribution, with a width which spreads like
$t^{1/2}$(i.e, $\nu=1/2$), therefore the variance grows linearly
with time. But the variance in the quantum walk on infinite line,
by contrast, grows quadratically with time, and the distribution
$\mid q_{k}(t)\mid^2$ has a complicated, oscillatory
form\cite{bca}. Now, by using the probability amplitudes of
Eq.(\ref{hp}) and Eq.(\ref{lp}) we evaluate the average of
different moments of  stratum number, i.e, $\langle k^q\rangle$
($q$-th moment, $q=1,2,3,...$ ), for quantum walk on infinite
Hermite,  Laguerre  and line graphs as follows:\\
\textbf{A.} Using the generating function Hermite polynomials,
the average $q$-moment of infinite Hermite graphs can be written
as
\begin{equation}
\langle k^q \rangle=\sum_{k=0}^{\infty}k^q\frac{(t^2)^k}{k!}
e^{-t^2}=e^{-t^2}(t^2\frac{d}{d(t^2)})^q\sum_{k=0}^{\infty}\frac{(t^2)^k}{k!}
=e^{-t^2}(t^2\frac{d}{d(t^2)})^qe^{t^2}.
\end{equation}
Above formula implies that $\langle k \rangle = t^2$ and $\langle
k^2 \rangle=t^4+t^2$, therefore for it standard deviation
$\sigma(t)$ we get $\sigma(t)=\sqrt{\langle k^2\rangle-\langle
k\rangle^2}=t$. Also, it is straightforward to see that for large
time, the $q$-moment has the following power dependence upon time
 \begin{equation}
\langle k^q \rangle\approx (t^{2})^{q}.
\end{equation}
 \textbf{B.} Similarly for infinite Laguerre graphs we have
 \begin{equation}
\langle k^q
\rangle=\sum_{k=0}^{\infty}k^q\frac{(\gamma+k)!}{k!\gamma!}
\frac{(at)^{2k}}{(1+(at)^2)^{k+\gamma+1}}
=\frac{1}{(1+a^2t^2)^{\gamma+1}}(y\frac{d}{dy})^q(\frac{1}{(1-y)^{\gamma+1}})|_{y=\frac{a^2t^2}{1+a^2t^2}},
\end{equation}
where it implies that
$$
 \langle k \rangle =(\gamma+1)a^2t^2, \;\;\;\
\langle k^2
\rangle=(\gamma+1)a^2t^2+(\gamma+1)(\gamma+2)(a^2t^2)^2
$$
\begin{equation}
\lim_{t\rightarrow \infty}(\sigma(t))=t^2, \;\;\;\;\;\
\lim_{t\rightarrow\infty} \langle k^q \rangle\approx (a^2t^2)^q.
\end{equation}
\textbf{C.} In infinite line graph, where the walker starts at
origin, we calculate the average moments of visiting number of
sites (therefore $k$ denotes the sites of graph where $k=0$
correspond to its origin). again using the well known the
generating Bessel function we can show that
$$
\langle k \rangle=\sum_{k=-\infty}^{\infty}2k\mid J_k(t)\mid^2=0,
$$
$$
\langle k^2 \rangle=\sum_{k=-\infty}^{\infty}2k^2\mid
J_k(t)\mid^2=t^2,
$$

\[
\lim_{t\rightarrow \infty}\langle k^q \rangle= = \left\{
\begin{array}{ll}
t^q & \mbox{for $q=$even}\\
0 & \mbox{otherwise.}
\end{array}
\right.
\]
\begin{equation}
\sigma(t)\approx t.
\end{equation}
 In the above examples the the average number of visiting strata (sites for infinite line), the
quantity, $\sigma(t)$ is the same as end-to-end distance. The
end-to-end distance for quantum walker on infinite Laguerre
graphs varies quadratically with time, i.e, $\langle
R^2\rangle\approx t^2$, while the end-to-end distance for walker
on infinite Hermite graphs is $\langle R^2\rangle\approx t$.
Thus, the quantum walk on infinite line and  Hermite graphs have
the same critical exponent, which is different from  infinite
Laguerre one. Therefore, quantum walk on infinite line and
Hermite graphs belong to the same universality class, which is
different from the universality class of quantum walk on infinite
Laguerre graph.
\subsection{Asymptotic behavior of the probability distribution
 in the large time limit}
In general we can not have an analytic expression for the
amplitudes of continuous time quantum walk on most of graphs ,
i.e, the integral   appearing in the Eq.(\ref{v4}) is difficult to
 evaluate, but one can approximate it for large time by using
 the method of stationary phase  explained above in section $4$.
 In order to see the importance of spectral distribution method,
 we obtain the asymptotic behavior of some finite
and infinite graphs in remaining part of this section.\\

 \textbf{7.2.1 The infinite graphs:}\\
  \textbf{7.2.1.1 Star lattice}\\
 \emph{Star lattice} is an $N$-fold star power $G^{\star
 N}$, where $G=(V,E)$ is the half line of integer, i.e., $V={0,1,2,...}$ and $i\sim
 j$ if and only if $\mid i-j\mid=1$, so its QD parameters  are $\omega_1=N, \omega_k=1, \alpha_k=0$ for $k=2,3,...$,
  and as it is shown in Ref.\cite{nob} it as the following spectral distribution
\begin{equation}
\mu(x)=\frac{1}{2\pi}\frac{N\sqrt{4-x^2}}{N^2-(N-1)x^2},
\;\;\;\;\;\ -2\leq x \leq 2.
\end{equation}
Therefore, its ground stratum amplitude $q_{0}(t)$ can be written
as
    \begin{equation}
 q_{0}(t)=\int_{-2}^{2}e^{-itx}\mu(dx)=
  \int_{-2}^{2}e^{-itx}(x)
 \frac{1}{2\pi}\frac{N\sqrt{4-x^2}}{N^2-(N-1)x^2}dx.
  \end{equation}
Now, its asymptotic form can be obtained by the method of
stationary phase  after making the change of variable
$x=2\cos(\theta)$, where  we get
\begin{equation}
q_{0}(t)\approx\frac{4N\Gamma(3/2)}{\pi(N-2)^2}\frac{1}{\sqrt{
t}}\cos(2t-\frac{3\pi}{4}),
\end{equation}
for $N\neq 2$, while, for $N=2$ the star lattices reduces to
infinite line graphs with  spectral distribution
$\mu(x)=\frac{1}{\pi}\frac{1}{\sqrt{4-x^2}}$,
 $-2\leq x \leq 2$ and asymptotic ground amplitude
 $q_{0}(t)=J_0(2t)\approx\frac{1}{\sqrt{\pi t}}\cos(2t-\frac{\pi}{4}) $.

\textbf{7.2.1.2 Two-dimensional comb lattice}

 As another example we consider two-dimensional comb lattice of Ref. \cite{ago},
with the following  spectral distribution
\begin{equation}
\mu(x)dx=\frac{1}{\pi}\frac{dx}{\sqrt{8-x^2}},   \;\;\;\;\
-2\sqrt{2}\leq x \leq 2\sqrt{2}
\end{equation}
and ground stratum amplitude
 $$
q_{0}(t)=\int_{-2\sqrt{2}}^{2\sqrt{2}}e^{-itx/4}\rho(x)dx=
\int_{-2\sqrt{2}}^{2\sqrt{2}}
\frac{1}{\pi}\frac{e^{-itx/4}dx}{\sqrt{8-x^2}}
$$
\begin{equation}
=\frac{1}{\pi}\int_{0}^{\pi}e^{-it\cos{\theta}/\sqrt{2}}d\theta.
\end{equation}
Again its asymptotic form can be obtained by the method of
stationary phase  after making the change of variable
$x=2\sqrt{2}\cos(\theta)$, where  we get
 \begin{equation}
 q_{0}(t)\sim \sqrt{\frac{2\sqrt{2}}{\pi t}}\cos(t/\sqrt{2}-\pi/4).
\end{equation}
\textbf{7.2.2 The finite graphs:} In order to obtain  the
asymptotic form  of finite graphs at large times , we need to use
Gauss quandrature  formula(\ref{ap}) to  approximate  finite
graphs with  the infinite ones and then using the method of
stationary phase we can obtain their asymptotic form.
 In the example under discussion asymptotic behavior of quantum
 walk on line and generalized to
asymptotic behavior quantum walk on finite graphs. For line
$q_{0}(t)$ is $J_{0}(t)$, then for large time $t$ we have
$q_{0}(t)\approx \sqrt{\frac{2}{t\pi}}\cos(t-\frac{\pi}{4})$. If
we use (\ref{ap}) and calculate $q_{0}(t)$ on finite graph line we
have
$q_{0}(t)=\frac{1}{n}\sum_{k=0}^{n-1}e^{-it\cos(\frac{2k+1}{2n}\pi)}$.
Now, we calculate numerically the  difference of amplitudes of
infinite line with finite graphs $\pi(n,t)$,  for large time $t$
$$
\pi(n,t)=\mid
J_{0}(t)-\frac{1}{n}\sum_{k=0}^{n-1}e^{-it\cos(\frac{2k+1}{2n}\pi)}\mid
$$
\begin{equation}
=\mid \sqrt{\frac{2}{t\pi}}\cos(t-\frac{\pi}{4})-
\frac{1}{n}\sum_{k=0}^{n-1}e^{-it\cos(\frac{2k+1}{2n}\pi)}\mid.
\end{equation}
The results are depicted in Fig.2 and Fig.3, where $\pi(n,t)$ is
in the limit $n$ from $500$ to $600$ and $t$ in the limited
$1000$, is limited to zero. Then, to study the behavior of
asymptotic quantum walk on
 finite graph  line, we study the behavior of asymptotic quantum walk on
 infinite graph line. Therefore to study the behavior of the asymptotic quantum walk
 on finite graphs, we can use arithmetic, approximate it
 with infinite graph, and by using the method of stationary phase,
  study the behavior of asymptotic quantum walk.
 \section{Conclusion}
 Using the spectral distribution associated with the adjacency
matrix, a new formalism for investigation of continuous-time
quantum walk on some  graphs is developed, where the orthogonal
polynomial together with their recursion relations play an
important role. As as by product, by using the Gaus quadrature
formula one can approximate infinite graph with finite one and
vice versa, which leads to the derivation of large time asymptotic
form of continuous-time quantum walk amplitudes of both finite
and infinite graph,  simply by using the method of stationary
phase. Even though the powerful method of spectral distribution
seems to work for some restricted kinds of graphs introduced in
Ref. \cite{nob}, but it is possible to generalize it to work for
most of graphs, particularly  for the graphs with strata dependent
$\kappa_{-}(y)$ and regular distant ones, which is under
investigation.

 \vspace{1cm}
\setcounter{section}{0}
 \setcounter{equation}{0}
 \renewcommand{\theequation}{A-\roman{equation}}
  {\Large{Appendix A}}\\
 In this appendix we prove the following lemma in connection
with the equality of continuous-time quantum walk amplitudes on
the vertices belonging to the same stratum.

\textbf{Lemma 1.} Let $q_{ik}(t)$ denote the amplitude of
observing the continuous-time quantum walker at
  vertex $i\in V_k$ at time $t$. Then for a class of  QD  graphs, the amplitude $q_{ik}$ is the same for all
 vertices of stratification $k$, for all $t$.\\
 \emph{Proof}.\\
 Let as take the Fourier transform of  unit vectors $\ket{k,i}$ of the stratum $V_k$:
\begin{equation}
\ket{\phi_{k,l}}=\frac{1}{\sqrt{\mid V_k\mid}}\sum_{i=0}^ {\mid
V_k\mid -1}\omega_{\mid V_k\mid}^{il}\ket{k,i},
\end{equation}
where $\omega_{\mid V_k\mid}=e^{\frac{2\pi i}{\mid V_k\mid}}$.
Now, acting the adjacency matrix $A$ on it according to the
formula (\ref{qd}), we obtain
\begin{equation}
A\ket{\phi_{k,0}}=\sqrt{\omega_{k+1}}\ket{\phi_{k+1,0}}+\alpha_{k+1}\ket{\phi_{k,0}}+
\sqrt{\omega_{k}}\ket{\phi_{k-1,0}}.
\end{equation}
Therefore, $\braket{\phi_{m,l}}{A\mid\phi_{k,0}}=0$, for $l\neq
0$, hence we have
$$
\braket{\phi_{k,l}}{e^{-iAt}\mid \phi_{0,l}}=\frac{1}{\sqrt{\mid
V_k\mid}}\sum_{i\in V_k}\omega_{\mid V_k\mid}^{il}p_{ik}(t)=0,
\;\;\;\;\; for, l\neq 0,
$$
\begin{equation}
q_k(t)=\frac{p_{0k}(t)+ p_{1k}(t)+p_{2k}(t)+\cdots+ p_{(\mid
V_k\mid)k}-1)(t)}{\mid V_k\mid} \;\;\;\;\; for, l= 0,
\end{equation}
or
\begin{equation}\label{l1}
F\left(\begin{array}{c} q_{0k}(t)\\ q_{1k}(t)\\q_{2k}(t)\\
\vdots\\ q_{(\mid V_k\mid-1) k}(t)
 \end{array}\right)
 =\left(\begin{array}{c}
 q_k(t)\\ 0 \\ 0 \\
 \vdots \\ 0
 \end{array}\right),
\end{equation}

where $\mid V_k\mid \times \mid V_k\mid$ the discrete Fourier
transformation matrix(DFT)  $F$ is defined as

\begin{equation}
 F = \frac{1}{\sqrt{\mid V_k\mid}}\left(\begin{array}{lllll}
     1 & 1 & 1 & \ldots & 1 \\
     1 & \omega & \omega^2 & \ldots & \omega^{\mid V_n\mid-1} \\
     1 & \omega^2 & \omega^4 & \ldots & \omega^{2(\mid V_n\mid-1)} \\
     \vdots & \vdots & \vdots &  & \vdots \\
     1 & \omega^{\mid V_n\mid-1} & \omega^{2(\mid V_n\mid-1)} & \ldots &
     \omega^{(\mid V_n\mid-1)^2}
    \end{array}\right).
\end{equation}
 Inverting  $F$ in  Eq.(\ref{l1}) we obtain
$q_{0k}(t)=q_{1k}(t)=\cdots=q_{(\mid V_k\mid-1) k
}(t)=\frac{q_{k}(t)}{\mid V_k\mid}$.

\vspace{1cm} \setcounter{section}{0}
 \setcounter{equation}{0}
 \renewcommand{\theequation}{B-\roman{equation}}
  {\Large{Appendix B}}\\
{\bf{ list of orthogonal polynomials connected with some finite and infinite graphs}}\\
In this appendix we give a list of orthogonal polynomials (in
addition to those mentioned above in section $6$), where after
appropriate one-dimensional, affine transformation(shift and
recalling), they can be  reduced to monic orthogonal polynomials
with recursion relation(\ref{op}), such that the corresponding
parameters $\omega_n$ and $\alpha_n$ can be connected with some
finite or infinite graphs of QD type.

\textbf{1.Charlier polynomials}\\
$Q_n(x)=a^n C_{n}^{(d)}(\frac{x-b}{a}),\;\;\;\
  \mu(x)=\sum_{j=0}\frac{e^{-d}d^x}{x!}\delta(x-j), \;\;\;\;\
  interval=(0,+\infty), \;\;\;\;\
   b=-ad,  \;\;\;\;\
  \omega_n=a^2nd, \;\;\;\;\
  \alpha_{n+1}=na.
  $\\
 Graph1.$\mid V_n\mid=\omega_1\omega_2\cdots\omega_n $(tree),\\
 Graph2.$\mid V_n\mid=n!a^2d^n$ and
$\kappa_{-}(y)=a.$\\
    \textbf{2.Meixner polynomials of the second kind}\\
  $Q_n(x)=a^nM_n(\frac{x-b}{a}),\;\;\;\
  \mu(x)=\mu(x;\delta,\eta)=(\Gamma (\eta/2))^{-2}\mid \Gamma
  (\frac{\eta+ix}{2})\mid^2 e^{-x\tan^{-1}(\delta)}, \;\;\;\;\
  interval=(-\infty,+\infty), \;\;\;\;\
   b=-a\eta \delta,  \;\;\;\;\
  \omega_n=a^2n(n+\eta+1)(\delta^2+1), \;\;\;\;\
  \alpha_{n+1}=2an\delta.
  $\\
 Graph1. $\mid V_n\mid=\omega_1\omega_2\cdots\omega_n $(tree),\\
 Graph2. $\mid V_n\mid=n!a^2(2+\eta)(3+\eta)\cdots (n+1+\eta) $ and
$\kappa_{-}(y)=a.$\\
 \textbf{3. Orthogonal polynomials
  related to Jacobi Elliptic
functions}\\
3.1\\ $Q_n(x)=a^nA_n(\frac{x-b}{a}), \;\;\
\mu(x)=\sum_{j=0}^{+\infty}\frac{kK^2}{\pi^2(2n)!(2n+1)!k^{2n}}\frac{(2j+1)q^{(2j+1)/2}}{1-q^{(2j+1)}}\delta(x-j),
\;\;\;\;\
  interval=(0,+\infty), \;\;\;\;\ b=-a(1+k^2), \;\;\
  \omega_n=4a^2n^2(4n^2-1)k^2, \;\;\;\;\
  \alpha_{n+1}=4an(n+1)(1+k^2)$.\\
 Graph1. $\mid V_n\mid=\omega_1\omega_2\cdots\omega_n $(tree),\\
Graph2. $\mid V_n\mid= 4^n (n!)^2a^2(1\times 3\times
15\times...\times(4n^2-1))k^{2n} $ and
$\kappa_{-}(y)=a$.\\
 3.2\\  $Q_n(x)=a^nB_n(\frac{x-b}{a}), \;\;\
 \mu(x)=\sum_{j=1}^{+\infty}\frac{k^2K^4}{2\pi^4(2n+1)!(2n+2)
 !k^{2n}}\frac{j^3q^j}{1-q^{2j}}\delta(x-j),
\;\;\;\;\
  interval=(1,+\infty), \;\;\;\;\  b=-4a(1+k^2),
\;\;\
  \omega_n=a^2 4n(n+1)(2n+1)^2 k^2, \;\;\;\;\
  \alpha_{n+1}=4an(n+2)(1+k^2)$.\\
 Graph1. $\mid V_n\mid=\omega_1\omega_2\cdots\omega_n $(tree), \\
Graph2. $\mid V_n\mid= 4^n n!(n+1)!a^2(3^2\times
5^2\times...\times(2n+1)^2)k^{2n} $ and
$\kappa_{-}(y)=a$.\\
3.3\\$Q_n(x)=C_n(x), \;\;\;\
\mu(x)=\sum_{j=-\infty}^{+\infty}\frac{\pi}{k
K(n)!^2k^{2[n/2]}}\frac{(q^{(2j+1)/2}}{1+q^{(2j+1)}}\delta(x-j),
\;\;\;\;\
  interval=(-\infty,+\infty), \;\;\;\;\  b=-4a(1+k^2),
\;\;\ \omega_{2n+1}=(2n+1)^2 , \;\;\ \omega_{2n}=(2n)^2k^2,
\;\;\;\;\
  \alpha_{n}=0$.\\
Graph1. $\mid V_n\mid=\omega_1\omega_2\cdots\omega_n$(tree).\\
3.4 \\
$Q_n(x)=D_n(x), \;\;\
\mu(x)=\sum_{j=-\infty}^{+\infty}\frac{\pi}{K(n)!^2k^{2[(n+1)/2]}}\frac{q^j}{1+q^{2j}}\delta(x-j),
\;\;\;\;\
  interval=(-\infty,+\infty), \omega_{2n+1}=(2n+1)^2k^2 , \;\;\
\omega_{2n}=(2n)^2,  \;\;\;\;\
  \alpha_{n}=0$.\\
Graph1. $\mid V_n\mid=\omega_1\omega_2\cdots\omega_n$(tree).\\
Such that, $j=0,\pm 1,\pm 2,\cdots,   \;\;\;\ K=\int_{0}^{\infty}(1-k^2\sin^2(\phi))^{-1/2}d\phi $\\
and $q$ is a certain constant related to $k$ which appears in the
theory of theta function.\\
3.5 \textbf{Carlitz polynomials}\\
3.5.1\\ $Q_n(x)=a^nF_n(\frac{x-b}{a}), \;\;\  b=-a, \;\;\
  \omega_n=4a^2n^2(2n-1)^2 k^2, \;\;\;\;\
  \alpha_{n+1}=4an(n(1+k^2)+1)$.\\
Graph1. $\mid V_n\mid=\omega_1\omega_2\cdots\omega_n$(tree),\\
Graph2. $\mid V_n\mid= 4^{n}a^2n!^2(1\times 3^2\times 5^2\times...\times (2n-1)^2)k^{2n}$ and $\kappa_{-}(y)=a$.\\
 3.5.2\\ $Q_n(x)=a^nG_n(\frac{x-b}{a}), \;\;\  b=-ak^2, \;\;\
  \omega_n=4a^2n^2(2n-1)^2 k^2, \;\;\;\;\
  \alpha_{n+1}=4an(n(1+k^2)+k^2)$.\\
Graph1.$\mid V_n\mid=\omega_1\omega_2\cdots\omega_n$(tree),\\
Graph2.$\mid V_n\mid= 4^{n}a^2n!^2(1\times 3^2\times 5^2\times...\times (2n-1)^2)k^{2n}$ and $\kappa_{-}(y)=a$.\\
 3.5.3\\ $Q_n(x)=a^nG_{n}^{\star}(\frac{x-b}{a}),\;\;\
b=-a(4+k^2), \;\;\  \omega_n=4a^2n^2(2n+1)^2 k^2, \;\;\;\;\
  \alpha_{n+1}=4an(n+1)(1+k^2)$.\\
Graph1. $\mid V_n\mid=\omega_1\omega_2\cdots\omega_n$(tree),\\
Graph2. $\mid V_n\mid= 4^{n}a^2n!^2(1\times 3^2\times 5^2\times...\times (2n-1)^2)k^{2n}$ and $\kappa_{-}(y)=a$.\\
 The explicit weight function for the subsection $(3.5)$ can be
obtained routinely from the weight function of the subsections
$(3.3)$ and
$(3.4)$.\\

\newpage
{\bf Figure Captions}

{\bf Figure-1:} Figure $G_4$.

{\bf Figure-2:} Shows $\pi(n,t)$ for the infinite line graphs with
the $n$-vertices finite graph (obtain from the infinite on via
Gauss quadrature rule) as a function $n$ at $t=1000$, where the
difference is almost negligible for $n\geq 518$.

{\bf Figure-3:} Figure$3$ show  $\pi(n,t)$ for the infinite
 line graph with the $n$-vertices finite graph as a
function of time and number of vertices $n$, where the difference
is almost disappear for $t\geq 1000$ and $n\geq 518$.


\begin{thebibliography}{99}
\bibitem{d}
P. Diaconis(1988), {\it Group Representation in Probability and
Statistics}, Hayward, California: Institute of Mathematical
Statistics.
\bibitem{ll}
L. Lovasz, {\it Random Walks on Graph: A Survey}, in
Combinatorics: Paul Erdos is Eighty, volume 2, edited by D.
Miklos, V.T. Sos, and T. Szonyi (Budapest: Janos Bolyai
Mathematical Society, 1996), 353-398.
\bibitem{rmr}
R. Motwani, and P. Raghvan(1995), {\it Randomized Algorithms},
Combridge University Press.
\bibitem{cn}
I. Chuang and M. Nielsen (2000), {\it Quantum Information and
Quantum Computation}, Cambridge University Press.
 \bibitem{diaconis}
P. Diaconis (1988), {\it Group Representations in Probability and
Statistics}, Institute of Mathematical Statistics.
\bibitem{fg}
E. Farhi and S. Gutmann (1998), {\it Quantum Computation and
Decision Trees}, Phys. Rev. A 58.
\bibitem{cfg}
E. Farhi, M. Childs, and S. Gutmann(2002), {\it An example of the
Difference between Quantum and Classical Random Walks}, Quantum
Information Processing, vol.1, p.35.
\bibitem{abnvw}
A. Ambainis, E. Bach, A. Nayak, A. Viswanath, and J. Watrous
(2001), {\it One-Dimensional Quantum Walks}, in Proceedings of
the  33rd ACM Annual Symposium on Theory Computing (ACM Press),
p. 60.
\bibitem{aakv}
D. Aharonov, A. Ambainis, J. Kempe, and U. Vazirani (2001), {\it
Quantum Walks on Graphs}, in Proceedings of the 33rd ACM Annual
Symposium on Theory Computing (ACM Press), p. 50.
\bibitem{mr}
C. Moore and A. Russell (2002), {\it Quantum Walks on the
Hypercube}, in Proceedings of the 6th Int. Workshop on
Randomization and Approximation in Computer Science (RANDOM'02).
\bibitem{k}
J. Kempe (2003), {\it Discrete Quantum Random Walks Hit
Exponentially Faster}, Proceedings of 7th International Workshop
on Randomization and Approximation Techniques in Computer Science
(RANDOM'03), p. 354-69.
\bibitem{fls}
R. Feynman, R. Leighton, and M. Sands(1965), {\it The Feynman
Lectures on Physics}, Volume 3, Addison-Wesley.
\bibitem{adz}
Y. Aharonov, L. Davidovich, and N.Zagury(1993), {\it Quantum
Randoms Walk}, Phy. Rev. lett 48, p.1687-1690.
\bibitem{ccdfgs}
A. Childs, E. Deotto, R. Cleve, E. Farhi, S. Gutmann, D. Spielman
(2003), {\it Exponential Algorithmic Speedup by Quantum Walk}, in
Proc. $35$th Ann. Symp. Theory of Computing (ACM Press), p. 59.
\bibitem{abtw}
A. Ahmadi, R. Belk, C. Tamon and C. Wendler(2003), {\it On Mixing
in Continuous-Time Quantum Walks on some Circulant Graphs},
Quantum Information and Computation, Vol. 3, No. 6, p. 611-618.
\bibitem{aaht}
W. Adamczak, K. Andrew, P. Hernberg, and C. Tamon(2003), {\it A
Note on Graphs Resistant to Quantum Uniform Mixing}, in
quant-ph/0308073.
\bibitem{gw}
H. Gerhardt, and J. Watrous(2003), {\it Continuous-Time Quantum
Walks on the Symmetric Group}, Proceedings of 7th International
Workshop on Randomization and Approximation Techniques in
Computer Science (RANDOM'03), p. 290-301.
\bibitem{obah}
A. Hora, and N. Obata(2002), {\it An Interacting Fock Space with
Periodic Jacobi Parameter Obtained from Regular Graphs in Large
Scale Limit}, to appear in: Quantum Information V, Hida, T., and
Saitoˆ , K., Ed., World Scientific, Singapore.
\bibitem{nob}
N. Obata(2004), {\it Quantum Probabilistic Approach to Spectral
Analysis of Star Graphs}, Interdisciplinary Information Sciences,
Vol. 10, No. 1, p. 41-52.
\bibitem{tsc}
T. S. Chihara, {\it An Introduction to Orthogonal Polynomials},
Gordon and Breach, Science Publishers Inc(1978).
\bibitem{st}
 J. A. Shohat, and J. D. Tamarkin, {\it The Problem of Moments, American Mathematical
 Society}, Providence, RI (1943).
\bibitem{obh}
 A. Hora, and N. Obata, {\it Quantum Decomposition and Quantum Central Limit
 Theorem}, in: Fundamental Problems in
Quantum Physics, Accardi, L., and Tasaki, S., Ed., p. 284–305,
World Scientific, Singapore(2003).
\bibitem{jf}
M. A. Jafarizadeh, and H. Fakhri(1997), {\it Supersymmetry and
Shape Invariance in Differential Equations of Mathematical
Physics}, Physics letters $A$, 230, p.164.
\bibitem{cbs}
Carl M. Bender, and Steven A. Orszag, {\it Advanced Mathematical
Methods for Scientists and Engineers}, International Series in
Pure and Applied Mathematics. McGraw-Hill, Inc., New York(1978).
\bibitem{nbrh}
N. Bleistein, and Richard A. Handelsman, {\it Asymptotic
Expansions of Integrals}, Holt, Rinehart and Winston, New York
(1975).
\bibitem{abt}
D. Ben-Avraham, E. Bollt, and C. Tamon(2004), {\it One-Dimensinal
Continuous-Time Quantum Walk}, Quantum Information Processing,
Vol. 3, p. 295-308.
\bibitem{koo}
N. Konno(2005), {\it Continuous-Time Quantum Walk on the Line}, in
quan-ph/0408140.
\bibitem{bca}
 T. A. Brun, H. A. Carteret, and A. Ambainis(2003), {\it Quantum Random
 Walks with Decoherent Coins }, Phys. Rev. A 67, 032304.
\bibitem{ago}
L. Accardi, ANIS B. Ghorbal, and N. Obata(2004), {\it Monotone
Independence, Comb Graphs and Bose-Einstein Condensation}, Infnite
Dimensional Analysis, Quantum Probability and Related Topics Vol.
7, No. 3, 419-435.
\end{thebibliography}
\end{document}